\documentclass[pra,aps, showpacs, twocolumn, a4paper,tightenlines,balancelastpage]{revtex4}
\usepackage{amsmath,amsfonts,amssymb,mathrsfs,bm}
\usepackage{verbatim,psfrag,times,CJK}
\usepackage{graphicx,graphics,color,epsfig}
\usepackage{flafter}

\usepackage{indentfirst}
\begin{document}

\title{Creation of pure multi-mode entangled states in a ring cavity}

\author{Gao-xiang \surname{Li}$^{a}$}
\email{gaox@phy.ccnu.edu.cn}

\author{Zbigniew \surname{Ficek}$^{b}$}
\affiliation{$^{a}$Department of Physics, Huazhong Normal University, Wuhan 430079, China\\
$^{b}$The National Centre for Mathematics and Physics, KACST, P.O. Box 6086, Riyadh 11442, Saudi Arabia}

\begin{abstract}
Practical schemes for creation of multi-mode squeezed (entangled) states of atomic ensembles located inside a high-$Q$ ring cavity are discussed. It is assumed that the cavity is composed of two degenerate mutually counter-propagating modes that can simultaneously couple to the atomic ensembles with the same coupling strengths. The ensembles are composed of ultra-cold atoms which are modeled as four-level systems driven by two laser fields, both co-propagating with one of the cavity directions. We illustrate a procedure that constructs multi-mode squeezed states from the vacuum by a unitary transformation associated with the collective dynamics of the atomic ensembles subjected to driving lasers of a suitably adjusted amplitudes and phases. The lasers pulses together with the cavity dissipation prepare the collective modes in a desired stationary squeezed state.
\end{abstract}

\pacs{03.67.Bg, 42.50.Dv, 42.50.Ex}

\maketitle

\section{Introduction}\label{int}

The problems of generation and practical applications of squeezed states of the electromagnetic field have been under a continuous interest since the early days of quantum optics~\cite{jmo,fds,df04}. Squeezed states, predicted in many phenomena of quantum optics and laser physics, are an example of non-classical states of the field which are characterized by a reduction of the quantum fluctuations in one of the field quadrature components below the usual vacuum level. The best description of how many different processes can exhibit squeezing was given by Krzysztof W\'odkiewicz, who comments "it is very easy and rather trivial to give many examples of quantum-mechanical states that lead to squeezed quantum fluctuations"~\cite{wo5}.
The ability to achieve a good quality squeezed light with significantly reduced fluctuations has not only resulted in the development of new areas of research but also provided a possibility of testing many fundamental ideas of quantum physics~\cite{ep35,ag82,w1,w2}.  Apart from the development of new theories of the interaction of quantum systems with the electromagnetic field, a number of  practical applications of squeezed light have been proposed and realized ranging from optical communications~\cite{hf85}, gravitational-wave detection~\cite{wc05} to newly emerging quantum technologies such as quantum information processing~\cite{ms06,yh07,pm09}, quantum teleportation~\cite{bk98,cw04}, quantum cryptography~\cite{h00,wd09} and quantum computing~\cite{yu08,gw09}.

Recent investigations of the continuous-variable quantum information have also generated a great deal of interest in squeezed states of light~\cite{kp03}. A particular attention has been paid to generation of multipartite continuous variable entangled states of atomic ensembles form multi-mode squeezed light produced, for example, in a non-linear process of parametric down conversion~\cite{bl05,jz03,at03,pf04}. The reason for using atomic ensembles is twofold. On the one hand, atomic ensembles are macroscopic systems that are easily created in the laboratory. On the other hand, the collective behavior of the atoms enables to achieve almost a perfect coupling of the ensembles to external squeezed fields without the need to achieve a strong field-single atom coupling.

Particularly interesting are schemes based on a quantum reservoir engineering, where squeezed states are generated inside the atomic ensembles by internal dynamical processes rather than being injected into the ensembles from external sources. The squeezed states are generated by a suitable driving that creates an effective multi-wave mixing processes to occur in the interaction of the fields with the atoms. For example, Rice and Carmichael~\cite{rc88} have shown that the fluorescence spectrum of single two-level atoms driven by a weak coherent laser field can be narrowed below the quantum limit giving a subnatural linewidth of the emitted field. The narrowing has been explained as resulting from squeezing in the fluctuations of the atomic dipole moments induced by the interaction of the atoms with the laser field. Narrowing of the spectral line below the quantum limit has also been predicted in the output field of a high-$Q$ microcavity engineered within a photonic crystal and containing two-level atoms driven by a strong laser field~\cite{tl08}. In this case, the spectral narrowing also results from squeezing, but the mechanism is different. The squeezing is produced by a non-linear mixing process resulting from a non-linear interaction of the cavity field with the dressed atom.

In a recent paper, Parkins {\it et al.}~\cite{ps06} have proposed a scheme for preparing single and two-mode squeezed states in atomic ensembles located inside a high-$Q$ ring cavity. The scheme, which is a generalization of the Guzman {\it et al.}~\cite{gr06} work to four-level atoms, is based on a suitable driving of the atomic ensembles with two external laser fields and coupling to a damped cavity mode that prepares the atoms in a pure squeezed (entangled) state. Similar schemes have been proposed to realize an effective Dicke model operating in the phase transition regime~\cite{de07}, to create a stationary subradiant state in an ultracold atomic gas~\cite{cb09}. This approach has recently been proposed as a practical scheme to prepare trapped and cooled ions in pure entangled vibrational states~\cite{li06} and to prepare four ensembles of hot atoms in pure entangled cluster states~\cite{lk09}.

The scheme proposed by Parkins {\it et al.}~\cite{ps06} assumes that atomic ensembles are effectively coupled to a single or two modes co-propagating with the driving fields.  In fact, a practical ring cavity is composed of two degenerate mutually counter-propagating modes that can simultaneously couple to the atomic ensembles with the same coupling strengths~\cite{krb03,na03,kl06}. With the counter propagating mode included the situation is somewhat more complicated, since the presence of the counter-propagating mode may result in a mode coupling which depends on atomic positions~\cite{gr00}.

\begin{figure}[thb]
   \begin{center}
   \begin{tabular}{c}
   \includegraphics[height=4.5cm,width=0.55\columnwidth]{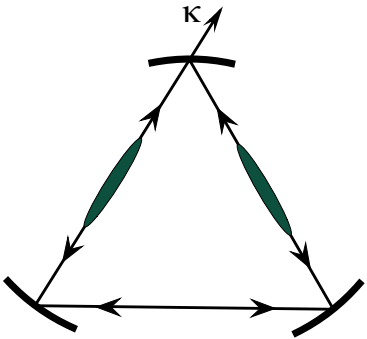}
   \end{tabular}
   \end{center}
   \caption[nsa]{ \label{fig1} A schematic diagram of a two-mode ring cavity containing two ensembles of cold atoms trapped along the cavity axis. Both cavity modes are damped with the same rate $\kappa$. The driving laser fields are injected through the cavity mirrors and co-propagate with one of the cavity modes.}
   \end{figure}

In this paper, we propose to generalize the Parkins {\it et al.} model to the case of a high-$Q$ ring cavity with two mutually  counter-propagating modes, as illustrated in Fig.~\ref{fig1}. In this two-mode configuration, the photon redistribution can occur between the counter-propagating modes which, as we shall see, will result in multi-mode squeezing. We propose a simple procedure to create multi-mode squeezed (entangled) states in the practical scheme involving cold atomic ensembles composed of four-level systems driven by sequences of laser pulses of specifically chosen Rabi frequencies and phases. The procedure constructs squeezed states from the vacuum by a unitary transformation associated with the dynamical process determined by the master equation of the system. With this more practical model, we illustrate how to create single and two-mode squeezed states in a single ensemble, and a four-mode squeezed state between two atomic ensembles.

Construction of pure squeezed states via unitary transformations, which we consider here,  is a fascinating research topic in which Krzysztof W\'odkiewicz was interested for many years. In a series of publications, he has explored similarities between the transformations and simple Lie algebras, and also investigated connections between squeezing and other quantum effects, such as the EPR paradox, photon antibunching and entanglement~\cite{wo1,wo2,wo3,wo4,wo6,wo7,wo8}.

\section{Effective Hamiltonian}\label{sec2}

\begin{figure}[thb]
   \begin{center}
   \begin{tabular}{c}
   \includegraphics[height=5cm,width=0.65\columnwidth]{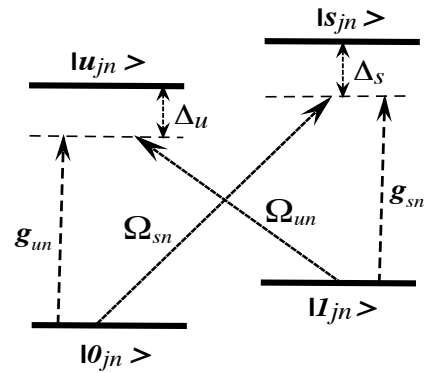}
   \end{tabular}
   \end{center}
   \caption[nsa]{ \label{fig2} Outline of the four-level system driven by two highly detuned laser fields and simultaneously coupled to cavity modes. The laser fields of the Rabi frequencies $\Omega_{un}$ and $\Omega_{sn}$ drive the atomic transitions $|1_{jn}\rangle\rightarrow |u_{jn}\rangle$ and $|0_{jn}\rangle\rightarrow |s_{jn}\rangle$, respectively. The cavity modes are coupled to atomic transitions $|1_{jn}\rangle\rightarrow |s_{jn}\rangle$ and $|0_{jn}\rangle\rightarrow |u_{jn}\rangle$ with the coupling strengths $g_{sn}$ and $g_{un}$, respectively.}
   \end{figure}

The physical system considered consists of two one-dimensional atomic ensembles trapped along the axis of a high-$Q$ ring cavity. Each ensemble is composed of $N$ identical multi-level atoms located at different positions $x_{jn}$ and modeled as four-level systems with two ground states $|0_{jn}\rangle, |1_{jn}\rangle$ and two excited states $|u_{jn}\rangle, |s_{jn}\rangle$, where the subscript $jn$ labels $j$th atom of the $n$th ensemble, see Fig.~\ref{fig2}. In practice such a four-level system could correspond to an $F=1\leftrightarrow F^{\prime}=1$ transition as occurs in $^{87}$Rb atoms. The cavity is composed of three mirrors that create two mutually counter-propagating modes, called clockwise and anti-clockwise  modes, to which the atoms are equally coupled. The cavity modes are degenerate in frequency, i.e.  $\omega_{+}=\omega_{-}=\omega_{c}$. Each ensemble is driven by two laser fields injected through the cavity mirrors and co-propagating with one of the cavity directions. The frequencies $\omega_{Lu}$ and $\omega_{Ls}$ of the laser fields are matched close to the cavity frequency, so that the wave numbers of the laser fields $k_{u}$ and $k_{s}$ are approximated by $k_u\approx k_s\approx k$. We assume that the laser fields and the cavity frequencies are far detuned from the atomic transition frequencies. This will allow us to adiabatically eliminate the upper states of the atoms and obtain an effective dispersive type interaction of the driving lasers and the cavity modes with the atomic transitions.

The Hamiltonian of the system, in the rotating-wave approximation, has the following form
\begin{eqnarray}
H=H_{0}+H_{AL}+H_{AC} ,\label{e1}
\end{eqnarray}
where
\begin{eqnarray}
H_{0} &=&  \hbar\omega_c (a_+^\dag a_{+} +a_{-}^{\dag} a_{-})
+ \hbar\sum\limits_{n=1}^{2}\sum\limits_{j=1}^N  \left\{\omega_u|u_{jn}\rangle\langle
u_{jn}|\right. \nonumber\\
&&+\left. \omega_s|s_{jn}\rangle\langle s_{jn}|+\omega_1|1_{jn}\rangle\langle
1_{jn}|\right\} \label{e2}
\end{eqnarray}
is the free Hamiltonian of the atomic ensembles and the cavity modes,
\begin{eqnarray}
H_{AL} &=&  \frac{1}{2}\hbar\sum\limits_{n=1}^{2}\sum\limits_{j=1}^N \left\{\Omega_{un}(x_{jn}) {\rm e}^{-i(\omega_{Lu}+\phi_{un})t}|u_{jn}\rangle\langle 1_{jn}|\right. \nonumber\\
&&+ \left. \Omega_{sn}(x_{jn}) {\rm e}^{-i(\omega_{Ls}+\phi_{sn})t}|s_{jn}\rangle\langle 0_{jn}| +{\rm H.c.}\right\}
\end{eqnarray}
is the interaction Hamiltonian between the atoms and the driving fields, and
\begin{eqnarray}
H_{AC} &=& \hbar\sum\limits_{n=1}^{2}\sum\limits_{j=1}^N\left\{
\left[g_{un}^{+}(x_{jn})a_{+} + g_{un}^{-}(x_{jn})a_{-}\right]|u_{jn}\rangle\langle0_{jn}|\right. \nonumber\\
&+&\left. \left[g_{sn}^{+}(x_{jn})a_{+} + g_{sn}^{-}(x_{jn})a_{-}\right]|s_{jn}\rangle\langle1_{jn}|+{\rm H.c.}\right\}
\end{eqnarray}
is the interaction Hamiltonian between the atoms and the two cavity modes.

Here, $a_{\pm}$ and $a^{\dagger}_{\pm}$ are the annihilation and creation operators associated with the two counter-propagating modes of the cavity; clockwise $(+)$ and anti-clockwise $(-)$ propagating modes, and $k$ is the wave number of the cavity modes. We have denoted the energies of the atomic levels by $\hbar\omega_{i} \, (i=1,u,s)$ and have set the energy of the ground state $|0_{jn}\rangle$ equal to zero. The parameters $\Omega_{un}(x_{jn})$ and $\Omega_{sn}(x_{jn})$ are the position dependent Rabi frequencies of the driving laser fields,  and $\phi_{un},\phi_{sn}$ are their phases. The coupling constants of the atomic transitions to the cavity fields, $g_{un}^{\pm}(x_{jn})$ and $g_{sn}^{\pm}(x_{jn})$, are also dependent on the atomic position. In what follows, we will use the plane traveling wave representation for the laser fields and the cavity modes, in which
\begin{eqnarray}
&&\Omega_{un}(x_{jn}) = \Omega_{un}{\rm e}^{ik_{u}x_{jn}} ,\quad \Omega_{sn}(x_{jn})
= \Omega_{sn}{\rm e}^{ik_{s}x_{jn}} ,\nonumber \\
&&g_{un}^{\pm}(x_{jn}) = g_{un}{\rm e}^{\pm ikx_{jn}} ,\quad g_{sn}^{\pm}(x_{jn})
= g_{sn}{\rm e}^{\pm ikx_{jn}} ,
\end{eqnarray}
where we have assumed that the coupling constants $g_{un}$ and~$g_{sn}$ are the same for the two cavity modes. This is acceptable if the modes have the same polarization and geometry, which is feasible with the current experiments~\cite{krb03,na03,kl06}.

We now make an unitary transformation and few standard approximations on the Hamiltonian (\ref{e1}) to eliminate the explicit time dependence and the excited states to neglect atomic spontaneous emission and to obtain an effective two-level Raman-coupled Hamiltonian. Then, we will make the transformation of the atomic operators into the field (bosonic) representation. These operations will be done in the following steps. In the first step, we make the unitary transformation $U =\exp(iH_{0}^{\prime}t/\hbar)$, with
\begin{eqnarray}
H_{0}^{\prime} &=& \hbar(\omega_{Ls}-\omega_1)(a_+^\dag a_++a_-^\dag a_-) \nonumber\\
&&+ \hbar\sum\limits_{n=1}^{2}\sum\limits_{j=1}^N\left\{ (\omega_u+\omega_1)|u_{jn}\rangle\langle u_{jn}|\right. \nonumber\\
&&\left. +\omega_{Ls}|s_{jn}\rangle\langle s_{jn}| + \omega_1|1_{jn}\rangle\langle 1_{jn}|\right\} ,
\end{eqnarray}
and assume that the laser frequencies satisfy the resonance condition
$\omega_{Ls}-\omega_{Lu}=2\omega_1$. This specifically chosen transformation allows us to eliminate the explicit time dependence of the Hamiltonian.

In the second step, we introduce detunings of the laser fields from the atomic transition frequencies
\begin{eqnarray}
\Delta_u=\omega_u-(\omega_{Lu}+\omega_1),\quad \Delta_s=\omega_s-\omega_{Ls} ,
\end{eqnarray}
and assume that the detunings are much larger than the Rabi frequencies and the atomic spontaneous emission rates.  This allows us to perform the standard adiabatic elimination of the atomic excited states and obtain an effective two-level Hamiltonian of the form
\begin{eqnarray}
\hat{H} &=& \left[\delta_{c}
+ \frac{1}{2}N\sum\limits_{n=1}^{2}\left( \frac{g_{un}^{2}}{\Delta_{un}} +\frac{g_{sn}^{2}}{\Delta_{sn}}\right)\right] \left(a_+^\dag a_+ +a_-^\dag a_{-}\right)\nonumber\\
&&+\sum\limits_{n=1}^{2}\left( \frac{g_{un}^{2}}{\Delta_{un}} -\frac{g_{sn}^{2}}{\Delta_{sn}}\right)
J^{(n)}_{z}\left(a_+^\dag a_+ +a_-^\dag a_{-}\right) \nonumber\\
&&+ \frac{1}{\sqrt{N}}\sum\limits_{n=1}^{2}\left\{\,
\beta_{un}{\rm e}^{-i\phi_{un}}\left(J_{0k}^{(n)}a_+^\dag +J_{2k}^{(n)}a_-^\dag \right)\right. \nonumber\\
&&+\left. \beta_{sn}{\rm e}^{-i\phi_{sn}}\left(J^{(n)\dagger}_{0k} a_+^\dag +J^{(n)\dagger}_{-2k}a_-^\dag\right)
+{\rm H.c.}\right\} \label{e7}
\end{eqnarray}
where $\delta_{c} =\omega_{c}-(\omega_{Ls}-\omega_{1})$ is the detuning of the cavity frequency from the Raman coupling resonance,
\begin{eqnarray}
&&J^{(n)}_{z} = \frac{1}{2}\sum\limits_{j=1}^N\left(|1_{jn}\rangle\langle 1_{jn}|-|0_{jn}\rangle\langle 0_{jn}|\right) = \sum\limits_{j=1}^N\sigma_{zj}^{(n)} ,\nonumber\\
&&J_{mk}^{(n)} = \sum\limits_{j=1}^{N}|0_{jn}\rangle\langle 1_{jn}|{\rm e}^{imkx_{jn}}
=\sum\limits_{j=1}^{N}\sigma^{(n)}_{j}{\rm e}^{imkx_{jn}}
\end{eqnarray}
are position dependent collective atomic operators, and
\begin{eqnarray}
\beta_{un}=\frac{\sqrt{N}\Omega_{un}g_{un}}{2\Delta_{un}} ,\quad
\beta_{sn}=\frac{\sqrt{N}\Omega_{sn}g_{sn}}{2\Delta_{sn}}
\end{eqnarray}
are the coupling strengths of the effective two-level system to the cavity modes. The three collective atomic operators, $J^{(n)}_{0k}, J^{(n)}_{2k}$ and $J^{(n)}_{-2k}$ arise naturally for the position dependent atomic transition operators and appear in a cavity with two mutually counter-propagating modes. In the case of a single-mode cavity, the Hamiltonian involves only the $J^{(n)}_{0k}$ operator~\cite{ps06,de07,cb09}. The operators $\sigma^{(n)}_{j}, \sigma^{(n)\dagger}_{j}$ and $\sigma_{zj}^{(n)}$ are the standard Pauli spin$-1/2$ operators, which satisfy the well known commutation relations $[\sigma_{j}^{(n)},\sigma_{\ell}^{(m)\dag}]=2\sigma_{zj}^{(n)}\delta_{j\ell}\delta_{mn}$. However, the collective operators $J^{(n)}_{mk}, J^{(n)\dagger}_{mk}$ and $J^{(n)}_{z}$ do not in general satisfy the angular momentum commutation relations. The reason is in the presence of the phase factors
$\exp(imkx_{jn})$, so that the commutation relations are satisfied only in the small sample limit of $kx_{jn}\ll 1$, at which $\exp(imkx_{jn})\approx 1$.

To avoid unessential complexity, we choose the frequencies of the driving lasers and the cavity field such that
\begin{eqnarray}
\frac{g_{un}^{2}}{\Delta_{un}} = \frac{g_{sn}^{2}}{\Delta_{sn}} , \qquad
\delta_{c} +\frac{Ng_{un}^{2}}{\Delta_{un}}= 0 ,
\end{eqnarray}
and find that after this simplification, the Hamiltonian (\ref{e7}) reduces to
\begin{eqnarray}
\hat{H} &=& \frac{1}{\sqrt{N}}\sum\limits_{n=1}^{2}\left\{\,
\beta_{un}{\rm e}^{-i\phi_{un}}\left(J_{0k}^{(n)}a_+^\dag +J_{2k}^{(n)}a_-^\dag \right)\right. \nonumber\\
&+&\left. \beta_{sn}{\rm e}^{-i\phi_{sn}}\left(J^{(n)\dagger}_{0k} a_+^\dag +J^{(n)\dagger}_{-2k}a_-^\dag\right)
+{\rm H.c.}\right\} .\label{e11}
\end{eqnarray}
This equation is in the form of a non-RWA Hamiltonian of two extended atomic ensembles independently coupled to two counter-propagating cavity modes. The parameters of the Hamiltonian are a function of the detunings and Rabi frequencies of the two highly detuned laser fields co-propagating with the clockwise cavity mode and thus could be controlled through the laser frequencies and intensities. In the small sample case of $kx_{jn}\ll 1$ and under the single-mode approximation, the Hamiltonian simplifies to the standard Dicke model~\cite{dic54,ft02}.

In the final step we shall reformulate the Hamiltonian (\ref{e11}) in terms of bosonic variables by adopting the Holstein-Primakoff representation of angular momentum operators~\cite{hp}. In this representation, the collective atomic operators, $J^{(n)\dagger}_{mk}, J^{(n)}_{mk}$ and $J^{(n)}_{z}$ are expressed in terms of annihilation and creation operators $C^{(n)}_{mk}$ and $C^{(n)\dagger}_{mk}$ of a single bosonic mode. Provided the atoms in each ensemble are initially prepared in their ground states $\{|0_{jn}\rangle\}$, and taking into account that due to large detunings of the driving fields, the
excitation probability of each atom is low during the laser-atom-cavity coupling, i.e., $\langle \sigma_{zj}^{(n)}\rangle \approx -1/2$, the collective atomic operators can be well approximated by
\begin{eqnarray}
J^{(n)}_{mk} = \sqrt{N} C^{(n)}_{mk} ,\quad J^{(n)}_{z} =  -\frac{N}{2} ,\label{e12}
\end{eqnarray}
where
\begin{eqnarray}
C^{(n)}_{mk} = \frac{1}{\sqrt{N}}\sum\limits_{j=1}^N c^{(n)}_{j}{\rm e}^{ imkx_{jn}} ,\quad m=0,\pm 2 ,\label{e13}
\end{eqnarray}
are collective bosonic operators with the operators $b^{(n)}_{j}$ and~$b^{(n)\dagger}_{j}$ obeying the standard bosonic commutation relation $[b^{(n)}_{j}, b^{(m)\dagger}_{\ell}]=\delta_{j\ell}\delta_{nm}$.

Note that the collective bosonic operators do not in general commute, i.e.
\begin{eqnarray}
\left[C^{(n)}_{mk},C^{(n^\prime)\dag}_{m^\prime k}\right] = \frac{1}{N}\sum\limits_{j=1}^N
\exp[i(m-m^\prime)kx_j] \delta_{nn^{\prime}} .\label{e14}
\end{eqnarray}
Hence, the modes are not independent of each other.

However, suppose that the atoms are in a chain with a constant separation $d$ between the adjacent atoms, which is much smaller than the cavity wavelength, i.e., $kd\ll 1$. Then the position of the $j$th atom can be expressed as $x_j=(j-1)d$, and the commutation relation (\ref{e14}) simplifies to
\begin{eqnarray}
\left[C^{(n)}_{mk},C^{(n^\prime)\dag}_{m^\prime k}\right]\approx
 \frac{\exp[i(m-m^\prime)kL]-1}{i(m-m^\prime)kL}\delta_{nn^{\prime}} ,\label{e15}
\end{eqnarray}
where $L=Nd$ is the length of the atomic ensemble.

If the length of the atomic medium is much larger than the cavity wavelength, $L\gg \lambda$, we obtain
\begin{eqnarray}
\left[C^{(n)}_{mk},C^{(n^\prime)\dag}_{m^\prime k}\right] \approx \delta_{m,m^\prime}\delta_{nn^{\prime}} ,\label{e16}
\end{eqnarray}
which shows that in the limit of small separations between the atoms and long atomic chains, the
the collective bosonic operators are orthogonal to each other. One can argue that at small separations between the atoms, assumed in the above derivation, the direct dipole-dipole interaction between the atoms is large and should be included into the calculations. However, we assume that the atoms are initially prepared in their ground state and owing the fact that the laser and the cavity mode frequencies are far from the resonance between the ground and excited states of the atoms, no transition dipole moments are induced between the atomic states. Thus, the direct dipole-dipole interaction between the atoms can be ignored.

Using the collective bosonic operators $C^{(n)}_{0k}$ and $C^{(n)}_{\pm 2k}$ we can rewrite the effective Hamiltonian (\ref{e11}) in the following form
\begin{eqnarray}
H_{e} &=& \sum\limits_{n=1}^{2}\left\{ \beta_{un} {\rm e}^{-i\phi_{un}}\left(\,C^{(n)}_{0k}+r_{0}C_{0k}^{(n)\dag}\right)a_+^\dag\right.
\nonumber\\
&+&\left. \beta_{un}{\rm e}^{-i\phi_{un}}\left(\,C^{(n)}_{2k}+r_{1}C^{(n)\dag}_{-2k}\right)a_-^\dag + {\rm H.c.}\right\} .\label{e17}
\end{eqnarray}
\begin{eqnarray}
r_{0,1} = \frac{\beta_{sn}}{\beta_{un}}{\rm e}^{-i(\phi_{sn}-\phi_{un})}
\end{eqnarray}
The important property of the bosonic representation is the fact that effective Hamiltonian of ensembles of cold atoms trapped inside a ring cavity can be expressed now as the interaction between the cavity modes and three orthogonal field modes; a collective mode $C^{(n)}_{0k}$ solely coupled to the cavity mode~$a_+$, which co-propagates with the driving lasers, and two modes $C^{(n)}_{\pm 2k}$ that are solely coupled to the cavity counter-propagating mode $a_{-}$.

The Hamiltonian (\ref{e17}) holds for the laser fields co-propagating with the clockwise mode only. Following the same procedure as above, we can easily show that in the case of the driving fields co-propagating with the anti-clockwise mode $a_-$,  the effective Hamiltonian takes the form
\begin{eqnarray}
&&H_{e} = \sum\limits_{n=1}^{2} \left\{ \left(\, \beta_{un}{\rm e}^{-i\phi_{un}}C^{(n)}_{0k}+\beta_{sn}{\rm e}^{-i\phi_{sn}}C_{0k}^{(n)\dag}\right)a_-^\dag\right. \nonumber\\
&&+\left. \left(\! \beta_{un}{\rm e}^{-i\phi_{un}}C^{(n)}_{-2k}\!+\!\beta_{sn}{\rm e}^{-i\phi_{sn}}C_{2k}^{(n)\dag}\right)a_+^\dag \!+\! {\rm H.c.}\right\} .\label{e18}
\end{eqnarray}
We see the complete symmetry between the two cases that reversing the direction of the propagation of the laser fields from clockwise to anti-clockwise is equivalent to the exchange of $a_{+}\leftrightarrow a_{-}$ and $k\rightarrow -k$ in the Hamiltonian (\ref{e17}).

Our objective is to prepare the atomic ensembles in a desired pure squeezed vacuum state. To achieve it, we consider the evolution of the system under the effective Hamiltonian (\ref{e17}) including also a possible loss of photons due to the damping of the cavity mode. This is the only damping which we will consider as we have already eliminated or minimized spontaneous emission by choosing large detunings of the driving lasers. The spontaneous emission rate due to off-resonant excitation is estimated at the rate $\gamma_{{\rm eff}}=\frac{1}{4}(\gamma/2\pi)(\Omega_{u(s)}/\Delta_{u(s)}^{2}$ that with typical experimental values of $\gamma = 6$\ MHz for a rubidium atom and $\Omega_{u(s)}/\Delta_{u(s)} = 0.005$ gives $\gamma_{{\rm eff}} \approx 40$\ Hz.
The estimated value for  $\gamma_{{\rm eff}}$ is significantly smaller than~$\kappa$ predicted for a cavity of the finesse $F= 1.7\times 10^{5}$.

With the cavity damping included, the properties of the system are determined by the density operator $\rho$ whose the time evolution is governed by the master equation
\begin{eqnarray}
\dot{\rho} = -i[H_{e} ,\rho]+{\cal L}_c\rho ,\label{e19}
\end{eqnarray}
where
\begin{equation}
{\cal L}_{c}\rho = \frac{1}{2}\kappa \sum\limits_{i=\pm}\left(2a_{i}\rho
a^\dag_{i}-a^\dag_{i} a_{i}\rho-\rho a^\dag_{i} a_{i}\right) ,\label{e20}
\end{equation}
is an operator representing the damping of the cavity field modes with the rate~$\kappa$.

In the following sections we will demonstrate how to generate on demand multi-mode squeezed vacuum states in ensembles of cold atoms located inside a two-mode ring cavity. We propose a simple procedure which constructs squeezed states from the vacuum by a unitary transformation associated with the realistic dynamical process determined by the master equation (\ref{e19}). In the procedure the atomic ensembles, initially in the ground state, are subjected to different sequences of laser pulses of specifically chosen Rabi frequencies $\Omega_{un}$ and $\Omega_{sn}$ and phases~$\phi_{un}$ and~$\phi_{sn}$. The laser pulses together with the cavity dissipation, that occurs with the rate $\kappa$, prepare the ensembles in a desired stationary state. We will assume in all our considerations that the initial ground state of the atomic ensembles corresponds to all the atoms being in their ground states~$|0_{jn}\rangle$.

\section{Creation of one and two-mode squeezed states in a single atomic ensemble}

First we illustrate a procedure which constructs single and two-mode squeezed states in a single $(n=1)$ ensemble of cold atoms located inside a two-mode ring cavity. The procedure constructs the squeezed states simply by acting on the vacuum with unitary operators called single and two-mode squeezed operators, respectively. The squeezed operators are defined as~\cite{cs85}
\begin{eqnarray}
S_0(\xi_0) &=& \exp\left[-\frac{1}{2}\left(\xi_0C_{0k}^{\dag 2}-\xi_{0}^{\ast}C_{0k}^2\right)\right] ,\nonumber\\
S_{\pm k}(\xi_1) &=& \exp\left(\xi_{1}^{\ast} C_{2k}C_{-2k} - \xi_{1}C_{2k}^\dag C_{-2k}^\dag\right) ,\label{e21}
\end{eqnarray}
where $\xi_{0}$ and $\xi_{1}$ are complex one and two-mode squeezing parameters, respectively. The squeezed operators can be easily associated with evolution operators for the effective
Hamiltonians~(\ref{e17}) and (\ref{e18}), which are already expressed in terms of the collective bosonic operators $C_{0k}^{(n)}$ and $C_{\pm 2k}^{(n)}$. The squeezing parameters are then given in terms of the coupling strengths $\beta_{un}$ and $\beta_{sn}$ and the phases~$\phi_{un}, \phi_{sn}$, so they can be adjusted and controlled by the driving laser fields.

The construction procedure is done in two steps. In the first step, we adjust the driving lasers to propagate in the clockwise direction, along the cavity mode $a_{+}$. In this case, the dynamics of the system are determined by the Hamiltonian (\ref{e17}). We then send series of laser pulses of phases $\phi_{u1}=\phi_{s1}=0$ and arbitrary Rabi frequencies~$\Omega_{u1}$ and $\Omega_{s1}$, but such that $\beta_{u1} >\beta_{s1}$.
With this choice of the parameters of the driving lasers and under the unitary squeezing transformation
\begin{eqnarray}
S_0(-\xi_0)S_{\pm k}(-\xi_{1})\rho S_0(\xi_0)S_{\pm k}(\xi_{1}) = \tilde{\rho} ,\label{e22}
\end{eqnarray}
with
\begin{eqnarray}
\xi_0=\xi_{1}=\frac{1}{2}\ln\left(\frac{\beta_{u1}+\beta_{s1}}{\beta_{u1}-\beta_{s1}}\right) ,\label{e23}
\end{eqnarray}
the master equation (\ref{e19}) becomes
\begin{equation}
\frac{d}{dt}\tilde{\rho}=-i[\tilde{H_e},\tilde{\rho}]+ {\cal L}_{c}\tilde{\rho} ,\label{e24}
\end{equation}
where
\begin{eqnarray}
\tilde{H_e} &=& S_0(-\xi_0)S_{\pm k}(-\xi_{1})H_eS_0(\xi_0)S_{\pm k}(\xi_{1})\nonumber\\
&=&\sqrt{\beta_{u1}^2-\beta_{s1}^2}\left(a_+^\dag C_{0k}+a_-^\dag C_{2k}+ {\rm H.c.}\right) .\label{e25}
\end{eqnarray}
It is seen that under the squeezing transformation, the Hamiltonian represents a simple system of two independent linear mixers, where the collective bosonic modes $C_{0k}$ and $C_{2k}$ linearly couple to the cavity modes $a_{+}$ and~$a_{-}$, respectively. The mode $C_{-2k}$ is decoupled from the cavity modes and therefore does not evolve. In other words, the state of the mode $C_{-2k}$ cannot be determined by the evolution operator for the Hamiltonian (\ref{e25}). The important property of the transformed system is that the master equation~(\ref{e24}) is fully soluble, i.e. all eigenvectors and eigenvalues can be obtained exactly. Hence, we can monitor the evolution of the bosonic modes towards their steady-state values. Since we are interested in the steady-state of the system, we confine our attention only to the eigenvalues of Eq.~(\ref{e24}), which are of the form
\begin{eqnarray}
\eta_\pm = -\frac{\kappa}{2}\pm\left[\left(\frac{\kappa}{2}\right)^2-\sqrt{\beta_{u1}^{2}-\beta_{s1}^{2}}\right]^\frac{1}{2} .\label{e26}
\end{eqnarray}
Evidently, both eigenvalues have negative real parts which means that the system subjected to a series of laser pulses up to a short time $t$ will then evolve (decay) to a stationary state that is a vacuum state. Thus, as a result of the interaction given by the Hamiltonian (\ref{e25}), and after a sufficiently long evolution time, the modes $a_{\pm}$, $C_{0k}$ and $C_{2k}$ will be found in the vacuum state, whereas the mode $C_{-2k}$ will remain in an undetermined state. The state of the mode $C_{-2k}$ will be determined in the next, second step of the preparation process.

In order to estimate the time scale for the system to reach the steady-state, we see from Eq.~(\ref{e26}) that as long as $\sqrt{\beta_{u1}^{2}-\beta_{s1}^{2}}>\kappa/2$, the time scale for the system  to reach the steady state is of order of $\sim 2/\kappa$. Thus, as a result of the cavity damping the system, after a sufficient long time, will definitely be found in the stationary state.

In summary of the first step of the preparation, we find that in the steady-state, the density matrix representing the state of the transformed system is in the factorized form
\begin{eqnarray}
\tilde{\rho}(\tau\sim 2/\kappa)= \tilde{\rho}_{v}\otimes\tilde{\rho}_{C_{-2k}} ,\label{e27}
\end{eqnarray}
where
\begin{eqnarray}
\tilde{\rho}_{v} =|0_{a_+},0_{a_-},0_{C_{0k}},0_{C_{2k}}\rangle
\langle 0_{a_+},0_{a_-},0_{C_{0k}},0_{C_{2k}}|  \label{e28}
\end{eqnarray}
is the density matrix of the four modes prepared in their vacuum states, and $\tilde{\rho}_{C_{-2k}}$ is the density matrix of the mode $C_{-2k}$ whose the state has not been determined in the first step of the procedure. The ket $|0_{a_+},0_{a_-},0_{C_{0k}},0_{C_{2k}}\rangle$ represents the state with zero photons in each of the modes.

Thus, we are left with the problem of the preparation of the remaining collective mode $C_{-2k}$ in a desired squeezed vacuum state. This is done in what we call the second step of the preparation, in which we first adjust the driving lasers to propagate along the anti-clockwise mode $a_{-}$. We then send series of pulses of frequencies, phases and amplitudes the same as in the above first stage. As a result of the coupling to the cavity mode $a_{-}$,  the interaction is now governed by the Hamiltonian (\ref{e18}), and therefore after the unitary squeezing transformation the Hamiltonian of the system takes the form
\begin{eqnarray}
\tilde{H_e} &=& S_0(-\xi_0)S_{\pm k}(-\xi_{1})H_eS_0(\xi_0)S_{\pm k}(\xi_{1})\nonumber\\
&=&\sqrt{\beta_{u1}^2-\beta_{s1}^2}\left(a_-^\dag C_{0k}+a_+^\dag C_{-2k} + {\rm H.c.}\right) .\label{e29}
\end{eqnarray}
As above in the case of the coupling to the cavity mode $a_{+}$, the Hamiltonian (\ref{e29}) describes a system of two independent linear mixers. Hence, the state of the system will evolve during the interaction   towards its stationary value, and after s suitably long time, $\sim 2/\kappa$, the transformed system will be found in the vacuum state.

Thus, after the second step of the preparation, the transformed system is found in the pure vacuum state determined by the density matrix of the form
\begin{equation}
\tilde{\rho}(\tau\sim 4/\kappa)=|\tilde{\Psi}\rangle\langle\tilde{\Psi} | ,\label{e30}
\end{equation}
where
\begin{eqnarray}
|\tilde{\Psi}\rangle &=&   S_0(-\xi_0)S_{\pm k}(-\xi_{1})|\Psi\rangle\nonumber\\
&=& |0_{a_{+}},0_{a_{-}},0_{C_{0k}},0_{C_{2k}},0_{C_{-2k}}\rangle \label{e31}
\end{eqnarray}
represents the vacuum state of the transformed system and the ket $|\Psi\rangle$ represents the final stationary state of the system.

If we now perform the inverse transformation from $\tilde{\rho}$ to $\rho$ or from $|\tilde{\Psi}\rangle$ to $|\Psi\rangle$, we find that in the steady-state the cavity modes are left in the vacuum state and the atomic ensemble is prepared in single and two-mode squeezed states. In other words, we find that the collective mode $C_{0k}$ is prepared in the one-mode squeezed vacuum state $S_0(\xi_0)|0_{C_{0k}}\rangle$, whereas the collective modes $C_{\pm2k}$ are in the two-mode squeezed vacuum state $S_{\pm k}(\xi_{1})|0_{C_{2k}},0_{C_{-2k}}\rangle$ associated with the superposition of two, position dependent counter-propagating modes. It is interesting to note that only a {\it single} set of the laser parameters, the Rabi frequencies and phases, was required to create the pure squeezed states.

The pure squeezed states are of particular importance in quantum physics as the density operator of the system prepared in the pure squeezed state cannot be factorized into the product of density operators of the individual modes. Thus, we may conclude that the atomic ensemble, after the interaction with the sequences of the laser pulses considered in this section is prepared in single and two-mode entangled states.

\section{Creation of a four-mode squeezed state with two atomic ensembles}

It is particularly interesting to generate entangled states between
two $(n=2)$  atomic ensembles. Since the cavity is composed of two
counter-propagating modes, the interaction between the collective
modes and the pair of cavity modes can produce squeezing of the
multi-mode variety, with a possibility to create a four-mode
squeezed state~\cite{yu08,ls87,msf07}. We will illustrate how it is
possible to create a four-mode squeezed state between two atomic
ensembles by a separate addressing of the collective modes with
sequences of laser pulses of suitably chosen Rabi frequencies and
phases.

Let us demonstrate a procedure which constructs the four collective
bosonic modes $C_{\pm2k}^{(n)}$ in the following pure four-mode
squeezed state
\begin{eqnarray}
|\Psi\rangle = S_{\pm k}(\xi)|0_{C_{2k}^{(1)}},0_{C_{-2k}^{(1)}},0_{C_{2k}^{(2)}},0_{C_{-2k}^{(2)}}\rangle ,\label{e32}
\end{eqnarray}
where the unitary operator has the form
\begin{eqnarray}
S_{\pm k}(\xi) &=&  \exp\left\{-\xi\left(C_{2k}^{(1)\dag}C_{-2k}^{(1)\dag}
+C_{-2k}^{(1)\dag}C_{2k}^{(2)\dag}\right.\right.\nonumber\\
&&+ \left.\left. C_{2k}^{(2)\dag}C_{-2k}^{(2)\dag} - {\rm
H.c.}\right)\right\} ,\label{e33}
\end{eqnarray}
and $\xi$ is the squeezing parameter which we take to be a real
number.   As in the above cases of one and two-mode squeezing, the
squeezing parameter can be adjusted and controlled by the driving
laser fields.

The action of the operator (\ref{e33}) on the vacuum state is to
produce  the four-mode squeezed state with correlations existing
between counter-propagating field modes of each and different
ensembles. Actually, the state $|\Psi\rangle$, after a suitable
local transformation can be related to a weighted four-mode
square-type cluster state, as shown by  Menicucci {\it et
al.}~\cite{msf07}. It has been demonstrated that the cluster state
(\ref{e32}) can be realized by engineering nonlinear type optical
interactions in quasi-phase-matched materials~\cite{s1},  such as
periodically poled KTiOPO4 (PPKTP) crystals~\cite{pfister}. As we
shall see, the state could be experimentally realized with two
ensembles of cold atoms placed inside a ring cavity.

Before introducing the procedure, we would like to point out that a
transformation of the density matrix  of the system with the unitary
operator (\ref{e33}) does not  lead to a density matrix describing a
simple system of independent linear mixers. Therefore, we first
transform the state $|\Psi\rangle$ to a new state $|\Phi\rangle =
T|\Psi\rangle$, with a unitary operator $T$ chosen such that it
transforms the field operators $C_{m2k}^{(n)}$ to new operators
$d_{m}^{(n)}=TC_{2mk}^{(n)}T^\dag$, which are linear combinations
involving the operators of different ensembles only, i.e.
\begin{eqnarray}
&& d_+^{\,(1)} = \frac{C_{2k}^{(1)}+\lambda C_{2k}^{(2)}}{\sqrt{1+\lambda^2}} ,\quad
d_+^{\,(2)}=\frac{\lambda C_{2k}^{(1)}- C_{2k}^{(2)}}{\sqrt{1+\lambda^2}},\nonumber\\
&& d_-^{\,(1)} = \frac{\lambda
C_{-2k}^{(2)}-C_{-2k}^{(1)}}{\sqrt{1+\lambda^2}} ,\quad
d_-^{\,(2)}=\frac{C_{-2k}^{(2)}+\lambda
C_{-2k}^{(1)}}{\sqrt{1+\lambda^2}} ,
\end{eqnarray}
where  $\lambda = (1+\sqrt{5})/2$.

With this carefully chosen transformation, we find that the state $|\Phi\rangle$ takes a form
\begin{eqnarray}
|\Phi\rangle &=& \exp\left[-\xi\left(\lambda C_{2k}^{(1)\dag}
C_{-2k}^{(2)\dag}-\frac{1}{\lambda} C_{2k}^{(2)\dag} C_{-2k}^{(1)\dag}\right) +{\rm H.c.}\right]\nonumber\\
&&\times \,|0_{C_{2k}^{(1)}},0_{C_{-2k}^{(1)}},0_{C_{2k}^{(2)}},0_{C_{-2k}^{(2)}}\rangle \nonumber\\
&=& S_{\pm k}^{(1)}(\lambda \xi)S_{\pm k}^{(2)}(-\xi/\lambda)|0_{C_{2k}^{(1)}},0_{C_{-2k}^{(1)}},0_{C_{2k}^{(2)}},0_{C_{-2k}^{(2)}}\rangle ,
\end{eqnarray}
where $S_{\pm k}^{(1)}(\lambda\xi)$ and $S_{\pm
k}^{(2)}(-\xi/\lambda)$ are  squeezing operators involving field
modes of the ensemble 1 and 2, respectively. As we shall see, the
advantage of working with the transformed state~$|\Phi\rangle$
rather than the state $|\Psi\rangle$ is that the density matrix
transformed with the operators $S_{\pm k}^{(1)}(\lambda \xi)$ and
$S_{\pm k}^{(2)}(-\xi/\lambda)$ describes a simple system of a
linear mixer of a cavity mode with one of the field modes.

We now proceed to perform the construction of the four-mode squeezed
state (\ref{e32}), which is done in four steps, each separately
addressing one of the collective modes. Since the collective modes
are orthogonal to each other, an arbitrary transformation performed
on one of the modes will not affect the remaining modes.

In the first step, we adjust the driving lasers to propagate in the
direction of  the clockwise mode~$a_{+}$. In this case, the dynamics
of the system are described by the Hamiltonian (\ref{e17}), and
after a unitary transformation of the density matrix
\begin{eqnarray}
S_{\pm k}^{(2)}(\xi/\lambda)S_{\pm k}^{(1)}(-\lambda\xi)T\rho T^\dag
S_{\pm k}^{(1)}(\lambda \xi)S_{\pm k}^{(2)}(-\xi/\lambda) = \rho_{1} ,\label{e36}
\end{eqnarray}
with the squeezing parameter
\begin{eqnarray}
\xi = \frac{1}{2\lambda}\ln\left(\frac{\beta_{u_1}+\beta_{s_2}}{\beta_{u_1}-\beta_{s_2}}\right) ,
\end{eqnarray}
and with the laser Rabi frequencies and phases such that
\begin{eqnarray}
&&\beta_{u_2} = \lambda\beta_{u_1} ,\quad \beta_{s_1}=\lambda\beta_{s_2} ,\nonumber \\
&&\phi_{u_n} = \phi_{s_n}=0 ,\quad n=1,2 ,\label{e38}
\end{eqnarray}
we find that the master equation of the transformed density matrix is of the form
\begin{equation}
\frac{d}{dt}\rho_{1} = -i[\tilde{H}_{1},\rho_{1}]+{\cal L}_{c} \rho_{1} ,\label{e39}
\end{equation}
where
\begin{eqnarray}
\tilde{H}_{1} &=& \left[\beta_{u_1}\left(C_{0k}^{(1)}+\lambda C_{0k}^{(2)}\right)
+\beta_{s_2}\left(\lambda C_{0k}^{(1)\dag}+C_{0k}^{(2)\dag}\right)\right]a_+^\dag \nonumber\\
&&+\sqrt{\left(1+\lambda^2\right)\left(\beta_{u_1}^2-\beta_{s_2}^2\right)}\, C_{2k}^{(1)}a_-^\dag
+{\rm H.c.} \label{e40}
\end{eqnarray}
We see that the choice of the laser parameters (\ref{e38}) results
in the collective mode $C_{2k}^{(1)}$, out of the four modes
involved in the  squeezing operator (\ref{e33}), being effectively
coupled to the cavity modes, the other modes are decoupled. It may
appear surprising that only one of the four collective modes is
effectively coupled to the cavity mode. However, this is merely a
consequence of the form of the effective interaction Hamiltonian
(\ref{e17}). It has an obvious advantage that we can separately
address each of the collective modes. One may also notice from
Eq.~(\ref{e40}) that apart form the collective mode $C_{2k}^{(1)}$,
the modes $C_{0k}^{(n)}$ are also coupled to one of the cavity
modes. Since the modes $C_{0k}^{(n)}$ are not involved in the
multi-mode squeezed state~(\ref{e32}) whose construction we are
interested in, we do not consider their evolution.

Fortunately, the master equation (\ref{e38}) is of a similar form
as Eq.~(\ref{e24}). Therefore, we can follow the same arguments as
before to conclude that the mode $C_{2k}^{(1)}$ will evolve in time
and definitely after a sufficiently long time, $\sim 2/\kappa$, the
mode will be found in a stationary vacuum state. The other three
modes will remain in undetermined states.

We now turn off the lasers propagating in the direction of the
clockwise  mode, and perform the second step, in which we first
adjust the driving lasers to propagate in the direction of  the
anti-clockwise mode~$a_{-}$.  In this case, we adjust the squeezing
parameters such that
\begin{eqnarray}
&&\beta_{u_1} = \lambda\beta_{u_2} ,\quad \beta_{s_2}=\lambda\beta_{s_1} ,\nonumber \\
&&\phi_{u_n} = \phi_{s_n}=0 ,\quad n=1,2 ,\label{e42}
\end{eqnarray}
to obtain the same value for the squeezing parameter now given by
\begin{eqnarray}
\xi = \frac{1}{2\lambda}\ln\left(\frac{\beta_{u_2}+\beta_{s_1}}{\beta_{u_2}-\beta_{s_1}}\right) .
\end{eqnarray}
For this choice of the parameters, and bearing in mind that the
dynamics of the system are now governed by the
Hamiltonian~(\ref{e18}), we find that after a unitary
transformation, Eq.~(\ref{e36}), the master equation of the
transformed density matrix takes the form
\begin{equation}
\frac{d}{dt}\rho_{2} = -i[\tilde{H}_{2},\rho_{2}]+{\cal L}_{c} \rho_{2} ,\label{e44}
\end{equation}
with
\begin{eqnarray}
\tilde{H}_{2} &=& \left[\beta_{u_2}\left(C_{0k}^{(1)}+\lambda C_{0k}^{(2)}\right)
+\beta_{s_1}\left(\lambda C_{0k}^{(1)\dag}+C_{0k}^{(2)\dag}\right)\right]a_{-}^\dag\nonumber\\
&&+\sqrt{\left(1+\lambda^2\right)\left(\beta_{u_2}^2-\beta_{s_1}^2\right)}\,
C_{-2k}^{(2)}a_{+}^\dag +{\rm H.c.}
\end{eqnarray}
We see that the choice of the laser parameters, Eq.~(\ref{e42}),
results in the  collective mode $C_{-2k}^{(2)}$ to be effectively
coupled to the cavity modes, with the other modes decoupled. The
dynamics of the mode are determined by the master equation
(\ref{e44}), which is of the similar form as Eq.~(\ref{e38}). Thus,
we can use the same arguments as in the previous step, and conclude
that after a sufficiently long time, $\sim 2/\kappa$, also the mode
$C_{-2k}^{(2)}$ will be found in a stationary vacuum state.

In the next step, we change the direction of propagation of the
driving lasers back to the direction of  the clockwise mode~$a_{+}$.
Similarly as in the above two steps, we first perform the unitary
transformation of the density matrix, Eq.~(\ref{e36}), with the
laser parameters chosen such that
\begin{eqnarray}
&&\beta_{u_1} = \lambda\beta_{u_2} ,\quad \beta_{s_2}=\lambda\beta_{s_1} ,\nonumber \\
&&\phi_{u_1} = \phi_{s_1}=0 ,\quad \phi_{u_2} = \phi_{s_2} = \pi ,\label{e47}
\end{eqnarray}
and
\begin{eqnarray}
\xi = \frac{\lambda}{2}\ln\left(\frac{\beta_{u_2}+\beta_{s_1}}{\beta_{u_2}-\beta_{s_1}}\right) .
\end{eqnarray}
With this choice of the laser parameters (\ref{e47}), we find that
now the  collective mode $C_{2k}^{(2)}$ is the only mode coupled to
the cavity modes and the dynamics of the mode follows the same
pattern as the modes considered in the above two steps. Thus, we may
conclude that after a sufficiently long time, also the collective
mode $C_{2k}^{(2)}$ will be found in a stationary vacuum state.

Finally, in the fourth step, we prepare the remaining
mode~$C_{-2k}^{(1)}$ in the vacuum state. To do this, we again
change the direction of the driving lasers to propagate in the
direction of the anti-clockwise mode and choose the laser parameters
such that
\begin{eqnarray}
&&\beta_{u_2} = \lambda\beta_{u_1} ,\quad \beta_{s_1}=\lambda\beta_{s_2} ,\nonumber \\
&&\phi_{u_1} = \phi_{s_1} = \pi ,\quad \phi_{u_2} = \phi_{s_2} = 0 ,\label{e49}
\end{eqnarray}
and
\begin{eqnarray}
\xi = \frac{\lambda}{2}\ln\left(\frac{\beta_{u_1}+\beta_{s_2}}{\beta_{u_1}-\beta_{s_2}}\right) .
\end{eqnarray}
Following the same procedure as in the previous three steps, one can
easily  show that the choice of the laser parameters,
Eq.~(\ref{e49}), results in a master equation for the transformed
density operator determined by the interaction Hamiltonian involving
the collective mode $C_{-2k}^{(1)}$. Thus, as a result of the
damping of the cavity modes, the collective mode will evolve towards
a vacuum state.

In this way, the final state of the transformed system is a four-mode vacuum state determined by the density operator
\begin{equation}
\tilde{\rho}(\tau\sim 8/\kappa)=|\tilde{\Phi}\rangle\langle\tilde{\Phi} | ,\label{e50}
\end{equation}
where
\begin{eqnarray}
|\tilde{\Phi}\rangle &=&   S_{\pm k}^{(1)}(-\lambda \xi)S_{\pm k}^{(2)}(\xi/\lambda)T|\Psi\rangle \nonumber\\
&=&  |0_{C_{2k}^{(1)}},0_{C_{-2k}^{(1)}},0_{C_{2k}^{(2)}},0_{C_{-2k}^{(2)}}\rangle \label{e51}
\end{eqnarray}
represents the vacuum state of the transformed system and the ket
$|\Psi\rangle$  represents the final stationary state of the system.

We see that in this practical scheme we can generate on demand an
arbitrary multi-mode squeezed state.  Generalizations of the scheme
to the case of three or more atomic ensembles located inside a ring
cavity are also possible and are trivial.

\section{Conclusions}\label{sec4}

We have proposed a procedure which constructs multi-mode squeezed
(entangled)  states in ensembles  of cold atoms located inside a
ring cavity. The procedure is referred to practical ring cavities
composed of three mirrors which create two counter-propagating modes
of degenerate frequencies. In contrast to the case of hot atomic
ensembles, where the one-mode approximation can be used to model the
dynamics of the atoms, the cold atomic ensembles are equally coupled
to both of the cavity modes. Using the the Holstein-Primakoff
representation of angular momentum operators, we have expressed
atomic systems in terms of orthogonal collective bosonic modes and
then have demonstrated how to construct multi-mode squeezed states
by a separate addressing of the collective modes. In particular, we
have shown how to construct one and two-mode squeezed states with a
single ensemble and four-mode squeezed states with two ensembles of
cold atoms. The procedure constructs squeezed states from the vacuum
by a unitary transformation associated with the collective dynamics
of the atomic ensembles subjected to driving lasers of a suitably
adjusted amplitudes and phases. The lasers prepare, with the help of
cavity dissipation, the collective modes in a desired stationary
squeezed state.

The procedures which construct squeezed (entangled) states by using
only  simple unitary transformations, the theme Krzysztof
W\'odkiewicz was interested for many years, offer further
interesting prospects for the study of multi-mode entanglement and
creation of entangled cluster states.

\section*{Acknowledgements}
This work is supported by the National Natural Science Foundation of
China (Grant Nos. 10674052 and 60878004), the Ministry of Education
under project NCET (Grant No. NCET-06-0671), SRFDP (under Grant No.
200805110002), and the National Basic Research Project of China
(2005 CB724508).







\end{document}